\documentclass[11pt]{article}
\usepackage{amsmath}
\usepackage{amsfonts}
\usepackage{amssymb}
\usepackage{epsfig}
\usepackage{times}
\newcommand{\p}{\partial}
\newcommand{\ol}{\overline}

\newcommand{\bs}{\boldsymbol}

\renewcommand{\l}{\newline\null}
\begin{document}
%
%
\begin{titlepage}
%
February 2003 (revised March 2003)\hfill PAR-LPTHE 03/09
\vskip 4.5cm
{\baselineskip 17pt
\begin{center}
{\bf $\bs{CP}$ NON-INVARIANCE FROM SCALING VIOLATIONS}
\end{center}
}
\vskip .5cm
\centerline{B. Machet
     \footnote[1]{Member of `Centre National de la Recherche Scientifique'}
     \footnote[2]{E-mail: machet@lpthe.jussieu.fr}
     }
\vskip 5mm
\centerline{{\em Laboratoire de Physique Th\'eorique et Hautes \'Energies}
     \footnote[3]{LPTHE tour 16, 1\raise 3pt \hbox{\tiny er} \'etage,
          Universit\'e P. et M. Curie, BP 126, 4 place Jussieu,
          F-75252 Paris Cedex 05 (France)}
}
\centerline{\em Universit\'es Pierre et Marie Curie (Paris 6) et Denis
Diderot (Paris 7)}
\centerline{\em Unit\'e associ\'ee au CNRS UMR 7589}
\vskip 1.5cm
{\bf Abstract:} 
In the framework of a renormalizable quantum field theory and taking the
example of neutral kaons, $CP$ violation is shown to be a dynamical
consequence of the anomalous scaling of the fields; its connection
to the mass splitting is exhibited. It is also shown to be
compatible with the normality condition $[M,M^\dagger]=0$ for the renormalized
mass matrix.
\smallskip

\vfill
\null\hfil\epsffile{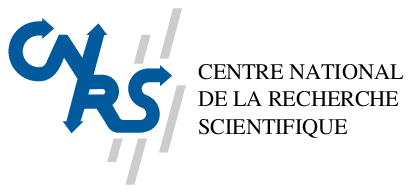}
\end{titlepage}
%

\section{Introduction}
\label{section:introduction}
%
$CP$ violation for pseudoscalar mesons is a well detected experimental fact
which is parametrized, in the quark framework, by a suitable mixing matrix
\cite{BrancoLavouraSilva}.
It stays a phenomenon {\em per se}, which has been lacking a
fundamental understanding, or, at least, a connection with other properties
of field theory.

In the framework of a renormalizable
quantum field theory,  indirect $CP$ violation
for neutral kaons (or similar systems) is shown in this letter to be
 directly connected to the
anomalous dimension of the neutral kaon fields, and, so, to scaling
violations.

The basic remark which straightforwardly leads to this result is that
the physical (different) masses $m_S$ and $m_L$ of the neutral mesons are not
the roots of the characteristic equation of a single constant
effective mass matrix \cite{Machet},
but are the poles of the full renormalized propagator of this binary system.
The components of each mass eigenvector are functions of the momentum $p$ which
must be evaluated at the solutions of the self-consistent equations
$p^2 = \lambda(p^2)$, the $\lambda(p^2)$'s being the eigenvalues of the full
renormalized self-energy matrix. 
Their evolution in the (small) interval $p^2\in[m_S^2,m_L^2]$ is accordingly
driven by renormalization group equations, with the consequence that
no more than one of them can be a $CP$ eigenstate as soon as they
scale with anomalous dimensions.

\section{Demonstration}
\label{section:dem}

\subsection{Physical masses and eigenstates}

Let $M^{(2)} (p^2)$, with dimension $[mass]^2$, be the full renormalized
self-energy $2\times 2$ matrix
for the system of neutral kaons in the $(K^0, \ol{K^0})$ basis; the
corresponding full inverse propagator writes 
\begin{equation}
L(p^2) = {\mathbb I}\,p^2 - M^{(2)}(p^2),\quad {\mathbb I}=diag(1,1).
\end{equation}
Accounting for unstable kaons requires $M^{(2)}(p^2)$ to be a complex
 non-hermitian matrix,
but we insist that, {\em for all $p^2$}, it is normal
\footnote{{\em i.e.} it commutes with its hermitian conjugate
$[M,M^\dagger]=0$.}
such that its right- and left-eigenvectors coincide \cite{Lowdin}
\footnote{No physically satisfying interpretation of a mismatch between the
two could ever be given; that our result fits in the case of a normal
matrix is thus welcome.}
; we demand furthermore that it satisfies the condition for $CPT$ invariance,
{\em i.e.} its two diagonal elements are identical \cite{BrancoLavouraSilva};
using the more convenient notation $z$ for the complex $p^2$, $M^{(2)}$
writes accordingly
\begin{equation}
M^{(2)}(z)= \left(\begin{array}{cc} a(z) &  b(z) \cr
                                    c(z) &  a(z) \end{array}\right),
\quad with\ b(z)\ \bar b(\bar z)= c(z)\ \bar c(\bar z).
\label{eq:M}
\end{equation}
The physical  masses
\footnote{They are complex as soon as the kaons are unstable
\cite{BrancoLavouraSilva}.}
$z_1\equiv m_S^2$ and $z_2\equiv m_L^2$ are solutions of
the self consistent equations $z-\lambda_\pm(z)=0$, $\lambda_\pm(z)$ being
the two solutions $\lambda_\pm(z) = a(z) \pm \sqrt{b(z)c(z)}$ of
$\det(M^{(2)}(z)-\lambda\ {\mathbb I})=0$, such that one has
\footnote{Of course, which is attributed to which is irrelevant.}
\begin{equation}
for\ K_S:\  z_1 = a(z_1) - \sqrt{b(z_1)c(z_1)},\quad
for\ K_L:\ z_2 = a(z_2) + \sqrt{b(z_2)c(z_2)}.
\label{eq:sols}
\end{equation}
The existence of other (spurious) solutions is irrelevant for what follows;
one only needs $z_1 \not = z_2$.

The components of the corresponding eigenvectors
$\left(\begin{array}{c}u_\pm(z) \cr v_\pm(z)\end{array}\right)$   of
$M^{(2)}(z)$ (and of $L(z)$) satisfy
\begin{equation}
\frac{v_\pm(z)}{u_\pm(z)} = \pm \sqrt{\frac{c(z)}{b(z)}}
=\pm\frac{z-a(z)}{b(z)},
\label{eq:ratios}
\end{equation}
and, so, in particular for the physical states
\footnote{The other eigenvectors $\left(\begin{array}{c}u_+(z_1)\cr
v_+(z_1)\end{array}\right)$ and $\left(\begin{array}{c}u_-(z_2)\cr
v_-(z_2)\end{array}\right)$ respectively of $M^{(2)}(z_1)$ and
$M^{(2)}(z_2)$ are spurious \cite{Machet}; they however still satisfy
(\ref{eq:ratios}).}
\begin{equation}
for\ K_S:\ \frac{v_-(z_1)}{u_-(z_1)} = - \sqrt{\frac{c(z_1)}{b(z_1)}} =
-\frac{z_1 - a(z_1)}{b(z_1)};\quad
for\ K_L:\ \frac{v_+(z_2)}{u_+(z_2)} = + \sqrt{\frac{c(z_2)}{b(z_2)}} =
+\frac{z_2 - a(z_2)}{b(z_2)}.
\label{eq:uv}
\end{equation}
The condition for the normality of $M^{(2)}$ (see at the right of (\ref{eq:M}))
entails in particular
\begin{equation}
\left | \frac{v_\pm(z)}{u_\pm(z)}\right |^2 =1.
\label{eq:norm}
\end{equation}

\subsection{Renormalization group equations}
\label{subsection:renorm}

We want to connect $v_+(z_2)/u_+(z_2)$ to $v_-(z_1)/u_-(z_1)$.

The elements of the renormalized full self-energy matrix
$M^{(2)}$, that we noted $a(z)$, $b(z)$ and $c(z)$ depend in reality not only
 on the external momentum $p$, but also of (dimensionless)
coupling constants $g_i$, masses $m_j$, (dimensionless) gauge parameters
$\xi_k$, and of a renormalization scale $\mu$; this is why, in the
following, we adopt instead the notation $a(p,g_i,m_j,a_k,\mu)$
{\em etc}, with $z=p^2$.
Defining $y_j = \frac{m_j}{\mu}$ and $t = \ln x$, $a(p,g_i,m_j,a_k,\mu)$
satisfies the renormalization group equation
(see appendix for the demonstration)
\begin{eqnarray}
&&\bigg( -\frac{\p}{\p t}
+ \sum_i \beta_i(g_i,y_j,\xi_k,\epsilon)g_i\frac{\p}{\p g_i}
- \sum_j \big(1+\gamma_j(g_i,y_j,\xi_k,\epsilon)\big)y_j\frac{\p}{\p y_j}
+ \sum_k \beta_{ak}(g_i,y_j,\xi_k,\epsilon)\frac{\p}{\p \xi_k}\cr
&& \hskip 1cm+ 2
- \frac{1}{2}\big(\gamma_{K^0}(g_i,y_j,\xi_k,\epsilon) + \gamma_{\ol{K^0}}(g_i,y_j,
\xi_k,\epsilon)\big) \bigg)\ a(e^t p, g_i,y_j,\xi_k,\mu) =0,
\label{eq:rg3}
\end{eqnarray}
where the $\beta_i$'s, $\gamma_j$'s, $\beta_{ak}$'s, and the anomalous
dimensions $\gamma_{K^0}$ and $\gamma_{\ol{K^0}}$ of the mesons
 are defined by
\begin{eqnarray}
&& \mu \frac{d g_i}{d \mu}= g_i\, \beta_i(g_i,y_j,\xi_k,\epsilon),\ 
   \frac{\mu}{m_j}\frac{d m_j}{d\mu}=-\gamma_j(g_i,y_j,\xi_k,\epsilon),\ 
   \mu\frac{d \xi_k}{d \mu} = \beta_{ak}(g_i,y_j,\xi_k,\epsilon),\cr
&& \frac{\mu}{Z_{K^0}(\mu,\cdots)}\frac{d Z_{K^0}(\mu,\cdots)}{d \mu}=
        \gamma_{K^0}(g_i,y_j,\xi_k,\epsilon),\ 
\frac{\mu}{Z_{\ol{K^0}}(\mu,\cdots)}\frac{d Z_{\ol{K^0}}(\mu,\cdots)}{d \mu}=
        \gamma_{\ol{K^0}}(g_i,y_j,\xi_k,\epsilon);\cr
&&
\label{eq:rg4}
\end{eqnarray}
$Z_{K^0}(\mu,\cdots)$ and $Z_{\ol{K^0}}(\mu,\cdots)$
are the renormalization constants for the wave functions of $K^0$ and
$\ol{K^0}$
\footnote{The renormalized and bare elements of $M^{(2)}$ are connected by
$a(\mu,\cdots) = \sqrt{Z_{K^0}(\mu,\cdots)}\sqrt{Z_{\ol{K^0}}(\mu,\cdots)}\,a_0,
b(\mu,\cdots) = Z_{K^0}(\mu,\cdots)
        \,b_0,\ 
c(\mu,\cdots)= Z_{\ol{K^0}}(\mu,\cdots)
        \,c_0$.}
.
The solution of (\ref{eq:rg3}) writes ($x=e^t$)
\begin{equation}
a(e^t p, g_i, y_j, \xi_k, \mu) = x^2 a(p, \ol{g_i}, \ol{y_j}, \ol{\xi_k}, \mu)
e^{-\int_0^t dt' \frac{1}{2}
\left(\gamma_{K^0}(\ol{g_i}(t',g_i),\ol{y_j}(t',g_i),\ol{\xi_k}(t',g_i),
\epsilon)
+\gamma_{\ol{K^0}}(\ol{g_i}(t',g_i),\ol{y_j}(t',g_i),\ol{\xi_k}(t',g_i),
\epsilon)\right)},
\label{eq:rg5}
\end{equation}
where the ``running'' $\ol{g_i}$, $\ol{y_j}$ and $\ol{\xi_k}$ are defined
by (at the extreme right are the initial conditions)
\begin{eqnarray}
&& \frac{d \ol{g_i}(t,g_j)}{dt}= \ol{g_i}\, \beta(\ol{g_i});\quad
\ol{g_i}(0,g_j)=g_i;\cr
&& \frac{d \ol{x_j}(t,g_i)}{dt}=-\big(1+\gamma_j(\ol{g_i})\big)\,\ol{x_j};
\quad \ol{x_j}(0,g_i)=x_j;\cr
&& \frac{d\ol \xi_k(t,g_i)}{dt}=\beta_{ak}(\ol{g_i});\quad
\ol{\xi_k}(0,g_i)=\xi_k.
\end{eqnarray}
The equations for $b$ and $c$ are similar to (\ref{eq:rg3}) with the only
replacement of $\frac{\gamma_{K^0} + \gamma_{\ol{K^0}}}{2}$ respectively by
$\gamma_{K^0}$ and $\gamma_{\ol{K^0}}$.

Let $x = p/p_1$. Since $z=p^2$ and $p_1^2=m_S^2$, (\ref{eq:rg5}) yields
(somewhat lightening the notations)
\begin{eqnarray}
&& a(xp_1,g_i,y_j,\xi_k,\mu) = \frac{z}{m_S^2}
        a(p_1,\ol{g_i},\ol{y_j},\ol{\xi_k})
e^{-\int_0 ^{(1/2)\ln(z/m_S^2)}dt'\frac{1}{2}
(\gamma_{K^0}(\ol{g_i}(t',g_i),\cdots)+
\gamma_{\ol{K^0}}(\ol{g_i}(t',g_i),\cdots) )},\cr
&& b(xp_1,g_i,y_j,\xi_k,\mu) = \frac{z}{m_S^2}
        b(p_1,\ol{g_i},\ol{y_j},\ol{\xi_k})
e^{-\int_0 ^{(1/2)\ln(z/m_S^2)}dt'
\gamma_{K^0}(\ol{g_i}(t',g_i),\cdots)
}.\cr
&&
\end{eqnarray}
In particular, for $z=z_2=m_L^2$, using (\ref{eq:uv}), one gets (with a
still more lightened notation)
\begin{equation}
\frac{v_+(z_2)}{u_+(z_2)}= \frac
{z_1-a(z_1)e^{-\int_0^{(1/2)\ln(m_L^2/m_S^2)}dt'
\frac{1}{2}(\gamma_{K^0}(t')+\gamma_{\ol{K^0}}(t') )}}
{b(z_1)e^{-\int_0^{(1/2)\ln(m_L^2/m_S^2)}dt' \gamma_{K^0}(t')
}}.
\end{equation}

\subsection{The case of a small mass splitting. Result}

Now, we use $m_L^2/m_S^2 \approx 1$ and develop the exponentials at the
first non-trivial order to get
\footnote{We use (\ref{eq:ratios}),  which yields at $z=z_1$:
$\frac{z_1-a(z_1)}{b(z_1)}= \frac{v_+(z_1)}{u_+(z_1)} = 
-\frac{ v_-(z_1)}{u_-(z_1)}$.}
\begin{equation}
\frac{v_+(z_2)}{u_+(z_2)} \approx \frac{v_+(z_1)}{u_+(z_1)}
\left(
1+\frac{a(z_1)}{z_1-a(z_1)}\delta_a + \delta_b
\right)
= - \frac{v_-(z_1)}{u_-(z_1)}\left(
1+\frac{a(z_1)}{z_1-a(z_1)}\delta_a + \delta_b
\right),
\end{equation}
with $\delta_a=\int_0^{(1/2)\ln(m_L^2/m_S^2)}dt' (1/2)(\gamma_{K^0}(t')
+\gamma_{\ol{K^0}}(t')$ and
$\delta_b = \int_0^{(1/2)\ln(m_L^2/m_S^2)}dt' \gamma_{K^0}(t')$.

In the small interval $[0,(1/2)\ln(m_L^2/m_S^2)]$, $\gamma_{K^0}$
and $\gamma_{\ol{K^0}}$ are expected to keep practically constant, such that
$\delta_a \approx (1/4)(\gamma_{K^0} +\gamma_{\ol{K^0}}) \ln(m_L^2/m_S^2)\approx
(\Delta m_K^2/4m_K^2)(\gamma_{K^0}+\gamma_{\ol{K^0}}) ,
\ \delta_b \approx (1/2)\gamma_{K^0}\ln(m_L^2/m_S^2))
\approx (\Delta m_K^2/2m_K^2)\gamma_{K^0}$, with
$\Delta m_K^2= m_L^2-m_S^2$.

Using again $\frac{\Delta m_K^2}{m_K^2}\ll 1$,
we can approximate, using (\ref{eq:sols}),
$a(z_1) \approx a(z_2) \approx (z_1+z_2)/2 \approx m_K^2$ and, consequently
$z_1 -a(z_1) \approx -\Delta m_K^2/2$, such that
%
$\frac{v_+(z_2)}{u_+(z_2)} \approx -\frac{v_-(z_1)}{u_-(z_1)}
\left(1 - \frac{\gamma_{K^0}+\gamma_{\ol{K^0}}}{2}
+ \frac{\Delta m_K^2}{2m_K^2}\gamma_{K^0}\right)$.
%
If the $K^0$ and $\ol{K^0}$ fields, related in quantum field theory
by hermitian conjugation,  scale accordingly with opposite anomalous dimensions
$\gamma_{K^0} + \gamma_{\ol{K^0}}=0$, the previous equation rewrites
($z_2=m_L^2$ and $z_1=m_S^2$)
\footnote{One gets the same result by studying the
scaling of $\sqrt{c(z)/b(z)}$ according to (\ref{eq:ratios}).}
\begin{equation}
\frac{v_+}{u_+}(m_L^2) \approx -\frac{v_-}{u_-}(m_S^2)
\left(1 + \frac{\Delta m_K^2}{2m_K^2} \gamma_{K^0}\right);
\label{eq:scal1}
\end{equation}
indirect $CP$ violation is consequently an unavoidable dynamical phenomenon
in renormalizable theories for mass split interacting neutral mesons with
anomalous scaling; 
suppose indeed that one of the mass eigenstate is a $CP$ eigenstate, {\em
i.e.} either $v_+(z_2)/u_+(z_2 = \pm 1$ or
$v_-(z_1)/u_-(z_1) = \pm 1$; in order to satisfy the
constraint (\ref{eq:norm}), the other ratio must acquire an imaginary part,
which forbids the corresponding state to be a $CP$ eigenstate.

\section{Conclusion}
\label{section:conclusion}

A general framework has been provided for the study of  $CP$ violation,
which is not limited to  neutral kaons and extends to other
species of particles;   all and only
those of the fundamental interactions which give rise to
violations of scaling are expected to contribute. Furthermore, by showing
that the renormalized mass matrix can be normal, all problems arising from the 
difference between right- and left-eigenstates have been wiped out.

%
\appendix
\section{Demonstration of equation (\ref{eq:rg3})}

The argumentation follows classical literature (see for example
 \cite{Coleman}).
Since $a(p, g_i, m_j,\xi_k,\mu)$ has dimension $[mass]^2$, it satisfies
\begin{equation}
a(xp, g_i, m_j,\xi_k,\mu)= \mu^2 \tilde a(x^2 \frac{z}{\mu^2},g_i,
\frac{m_j}{\mu},\xi_k),
\end{equation}
where $\tilde a$ is an homogeneous function of degree $0$ in $x$, $m_j$, $\mu$;
similar equations for $b$ and $c$ define the homogeneous functions
$\tilde b$ and $\tilde c$.
Euler's theorem  applied to $\tilde a$ yields
\begin{equation}
\left(x\frac{\p}{\p x} + \sum_j m_j\frac{\p}{\p {m_j}} + \mu\frac{\p}{\p\mu}
-2\right) a(xp, g_i, m_j,\xi_k,\mu) =0,
\label{eq:rg1}
\end{equation}
and similar equations for $b$ and $c$.
On the other side, since the bare quantities $a_0$, $b_0$ and $c_0$ do not
depend on $\mu$
\begin{equation}
\mu\frac{\p}{\p\mu} a_0(p,g_{i0},m_{j0},a_{k0},\epsilon)=0,\  etc
\end{equation}
where we have supposed the regularization done by going to $4-\epsilon$
dimensions.
This yields (see notations and footnote in subsection \ref{subsection:renorm})
\begin{eqnarray}
&&\left( \mu\frac{\p}{\p\mu} +
\sum_i \mu\frac{dg_i}{d \mu}\frac{\p}{\p g_i} + 
\sum_j \frac{\mu}{m_j}\frac{d m_j}{d \mu}m_j\frac{\p}{\p m_j} + 
\sum_k \mu\frac{d\xi_k}{d \mu}\frac{\p}{\p \xi_k} \right)
a(p,g_i,m_j,\xi_k,\mu)\cr
&&
= \frac{1}{2}\bigg(\frac{\mu}{Z_{K^0}(\mu,\cdots)}
\frac{dZ_{K^0}(\mu,\cdots)}{d\mu}a(p,g_i,m_j,\xi_k,\mu)
+ \frac{\mu}{Z_{\ol{K^0}}(\mu,\cdots)}
\frac{dZ_{\ol{K^0}}(\mu,\cdots)}{d\mu}a(p,g_i,m_j,\xi_k,\mu)\bigg), \cr
&&
\label{eq:rg2}
\end{eqnarray}
which rewrites
\begin{equation}
\left( \mu\frac{\p}{\p\mu} +
\sum_i \beta_i g_i\frac{\p}{\p g_i}
-\sum_j \gamma_j m_j \frac{\p}{\p m_j}
+ \sum_k \beta_{ak}\frac{\p}{\p \xi_k}\right)
a(p,g_i,m_j,\xi_k,\mu)
= \frac{\gamma_{K^0}+ \gamma_{\ol{K^0}}}{2} a(p,g_i,m_j,\xi_k,\mu).
\label{eq:rg22} 
\end{equation}
(\ref{eq:rg22}) also applies to $ a(xp,\cdots,\mu)$; 
replacing there $\mu\, {\p a(xp,\cdots,\mu)}/{\p\mu}$ by its value extracted
from (\ref{eq:rg1}) gives equation (\ref{eq:rg3}).

%
%
%
\begin{em}

\end{em}


\end{document}